# Spin and Charge Texture around In-Plane Charge Centers in the $CuO_2$ planes

Stephan Haas(1), Fu-Chun Zhang(1,2), Frederic Mila(3), and T.M. Rice(1)
(1) *Theoretische Physik, ETH-Hönggerberg, CH-8093 Zurich, Switzerland*
(2) *Department of Physics, University of Cincinnati, Cincinnati, OH 45221*
(3) *Laboratoire de Physique Quantique, Université Paul Sabatier, 31062 Toulouse, France*

Recent experiments on $La_2Cu_{1-x}Li_xO_4$ show that although the doped holes remain localized near the substitutional Li impurities, magnetic order is rapidly suppressed. An examination of the spin texture around a bound hole in a $CuO_2$ plane shows that the formation of a skyrmion is favored in a wide range of parameters, as was previously proposed in the context of Sr doping. The spin texture may be observable by elastic diffuse neutron scattering, and may also have a considerable effect on NMR lineshapes.

Although it has been known for some time that only a very small concentration of holes doped into the antiferromagnetic (AF) state of the cuprates suffices to suppress long range AF order, the close proximity to the metal insulator transition makes it difficult to distinguish clearly between the length scales of charge and magnetic disturbances. Recently, experiments on the $La_2Cu_{1-x}Li_xO_4$ system where $Li^+$ substitutes for $Cu^{2+}$ in the $CuO_2$-planes show a similar rapid suppression of long range AF order but without a transition to a conducting state. [1] In this system clearly the hole charge introduced by the $Li^+$-substitution is well localized, whereas the magnetic perturbation must be long ranged. Several authors have in the past proposed that bound holes introduce skyrmion textures into the AF ordered state. [2–4] In this letter, we model the Li-substitution and examine this possibility in greater detail in the limit of dilute Li-concentration. Our conclusion is that skyrmion textures which are known to destroy long range magnetic order [5] accompany holes bound to charge centers, and we propose that the skyrmion textures may be observable in neutron scattering and NMR experiments.

The substitution of $Li^+$ for $Cu^{2+}$ in a cuprate magnetic insulator such as $La_2CuO_4$ has two effects. First, a $Li^+$ ion is a net negatively charged center, and secondly each $Li^+$ ion is accompanied by a doped hole. This hole in turn is attracted to the $Li^+$ site by a Coulomb potential, but it is not allowed to hop onto the impurity site. Additionally, there is a change of the singlet $Cu^{3+}$ state of the hole on neighboring Cu-sites due to a modification of the pd-hybridization on these sites.

To gain an intuitive understanding of the effect of a partially localized hole on the local magnetic moment at neighboring $Cu^{2+}$ sites in the plane, let us first consider the atomic limit. To zeroth order, a static hole which is attracted to a $Li^+$ impurity (located at site $\mathbf{0}$) by a nearest neighbor Coulomb interaction ($-\tilde{V}\sum_{\pm \mathbf{e_x},\mathbf{e_y}} n_{\mathbf{0}} n^h_{\pm \mathbf{e_x},\mathbf{e_y}}$, where $\mathbf{e_x}$ and $\mathbf{e_y}$ are unit vectors linking the impurity site with neighboring oxygen sites) will reside on one of the $CuO_2$ squares in the innermost "shells" - consisting of the 4 $CuO_2$ adjacent to the impurity ion (Fig. 1). The energy of the resulting singlet is given by

$$E_1 = \frac{E_B}{4}[\alpha + \sqrt{3}\sqrt{1-\alpha^2}]^2 + \alpha^2\tilde{V}, \quad (1)$$
$$\alpha^2 = \frac{1}{2}[1 + \frac{2\tilde{V} - E_B}{\sqrt{(2\tilde{V}-E_B)^2 + 3E_B^2}}],$$

where $E_B$ is the binding energy of the singlet, while the factor $\alpha$ takes into account the anisotropy of the singlet due to its proximity to the impurity. Specifically, the distortion of the singlet structure ($\phi_i = \frac{1}{\sqrt{2}}(d^\dagger_\uparrow p^\dagger_{i\downarrow} - d^\dagger_\downarrow p^\dagger_{i\uparrow})$) [7] around $Cu^{3+}$ sites adjacent to the impurity leads to a distorted singlet of the form $\Phi = \frac{1}{2}[\alpha\phi_1 + \frac{1}{\sqrt{3}}\sqrt{1-\alpha^2}(-\phi_2 + \phi_3 - \phi_4)]$, where the index 1 refers to the O-site adjacent to the impurity.

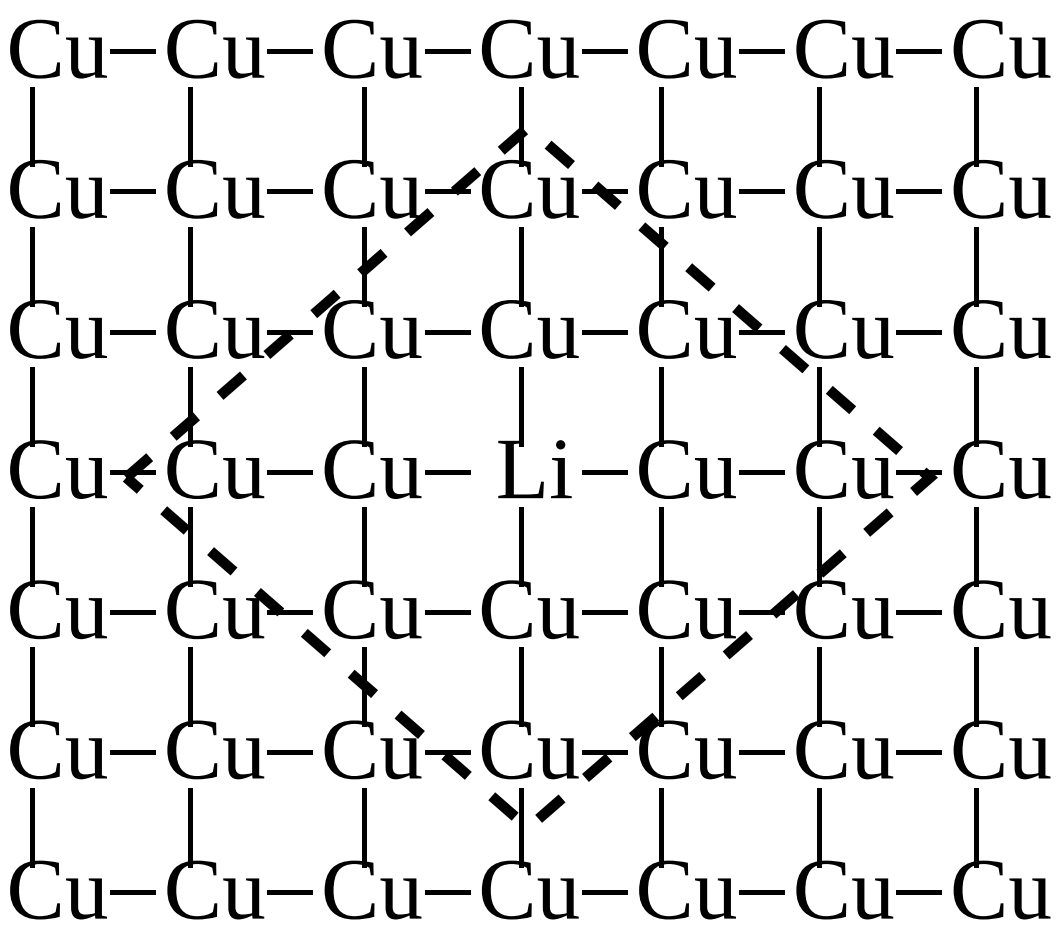

FIG. 1. Schematic plot of the 13-site cluster consisting out of 3 Cu-shells centered around a Li impurity. O-sites on the Cu-Cu and the Cu-Li links are not shown.

Introducing a hopping term, we find within second order perturbation theory about the atomic limit that the average hole occupancy on a Cu-site in the $i^{th}$ shell around the impurity site is given by

$$< n_i^h >= \frac{1}{4} \cdot \frac{\delta_{i,1}(E_1 - E_B)^2 + a_i t^2}{(E_1 - E_B)^2 + 5t^2}, \tag{2}$$

where t is the nearest neighbor hopping integral between the Cu-sites, and $a_1 = 0$, $a_2 = 4$, $a_3 = 1$. The corresponding magnetic moment in the $i^{th}$ shell is then given by $< S_i >= \frac{1}{2}(1- < n_i^h >)$. [8] Thus, for a strong enough Coulomb attraction, the hole will be constrained almost entirely to the innermost Cu-shell around the impurity yielding an average magnetic moment of $< S_1 >\simeq \frac{3}{8}$, i.e. a reduction of 25%. In addition, corrections due to zero-point fluctuations are expected when exchange processes among the spins are included.

Let us now turn to the question of the spin background created by a dilute concentration of holes, which are partially localized around impurity ions, in an otherwise AF background. While the motion of a hole about an impurity ion has strong quantum character, tending to spin-polarize the region around the impurity, the area far from the impurity may be treated in a semi-classical manner. [2] It is then natural to assume that the spin background will compromise between the (partially) polarized area around the impurity ion and the AF long-range order which is expected at infinite distance from the impurity by favoring a staggered skyrmion configuration. [2–4] In this sense, the "quantum" region where the hole moves rapidly around the impurity imposes a boundary condition on the quasi-classical spin background, which in turn may be described by an O(3) continuum model that is renormalized by quantum fluctuations. [4]

As first noted by Belavin and Polyakov, [5] skyrmions are a metastable solution of the O(3) non-linear $\sigma$-model with Hamiltonian $H_\sigma = \frac{1}{2}\rho_s^2 \int d^2r(\nabla \mathbf{n})^2$. Their solution for the two-dimensional isotropic ferromagnet can be extended to the AF case, [2,3] which - for a given choice of sign of the topological charge - has the form

$$\begin{aligned} n_x &= (-1)^{x+y}\frac{2\lambda x}{r^2+\lambda^2}, \\ n_y &= (-1)^{x+y}\frac{2\lambda y}{r^2+\lambda^2}, \\ n_z &= (-1)^{x+y}\frac{r^2-\lambda^2}{r^2+\lambda^2}, \end{aligned} \tag{3}$$

where $\lambda$ is the core size of the skyrmion, and the alternating sign corresponds to the two sublattices. We will demonstrate that such a spin texture is indeed stable in the presence of holes that are partially localized around defects in the plane.

In the following, the region around the impurity is treated exactly, i.e. by exactly diagonalizing a cluster of the kind shown in Fig. 1, while the semi-classical spin background enters the finite system calculation as an outer boundary condition through an exchange field, $\mathbf{JS(r)} = \frac{J}{2}\mathbf{n}(\mathbf{r}, \lambda)$. Then the free energy of the whole system (finite cluster + surrounding field) is minimized with respect to $\lambda$ to deduce the optimal core size $\lambda_{opt}$ of the skyrmion field. A minimum at $\lambda_{opt} = \infty$ or $\lambda_{opt} = 0$ would imply that the system favors AF alignment, and spin currents introduced by defects would be irrelevant. On the other hand, a finite $\lambda_{opt}$ corresponds to a stable skyrmion spin texture.

We study the planar t-J model with one hole on the 13-site cluster shown in Fig. 1, with boundary conditions imposed by equation (3). The in-plane impurity site, located at the origin $\mathbf{0}$ of the cluster, attracts the hole via a nearest-neighbor Coulomb potential ($|V| \approx 0.5t$ for Li). The corresponding Hamiltonian is given by

$$\begin{aligned} H = &- t \sum_{<i,j>\neq \mathbf{0},\sigma} (\tilde{c}^\dagger_{i\sigma}\tilde{c}_{j\sigma} + h.c.) \\ &+ J \sum_{<i,j>\neq \mathbf{0}} (\mathbf{S}_i \cdot \mathbf{S}_j - \frac{1}{4}n_i n_j) - V \sum_{<\mathbf{0},\pm\mathbf{x},\mathbf{y}>} n_{\mathbf{0}} n^h_{\pm\mathbf{x},\mathbf{y}}, \end{aligned} \tag{4}$$

where the $\tilde{c}$-operators are hole operators acting on non-doubly occupied states, J is the exchange integral, and t is the hopping amplitude. This system is not translationally invariant because of the defect. Furthermore, the twist components $(S_x, S_y)$ of the applied skyrmion field at the boundary of the cluster create non-vanishing matrix elements between different $S^z_{tot}$ subspaces of the enclosed system. In the absence of the outer spin field, the ground state of the impurity-containing cluster is doubly degenerate ($S^z_{tot} = \pm\frac{1}{2}$), and has a non-vanishing associated current, $j^\alpha = it[(\tilde{c}^\dagger_{i\sigma}\hat{\tau}^\alpha_{\sigma,\sigma'}\tilde{c}_{i+\hat{\mathbf{1}},\sigma'} - \tilde{c}^\dagger_{i+\hat{\mathbf{1}}\sigma}\hat{\tau}^\alpha_{\sigma,\sigma'}\tilde{c}_{i,\sigma'}]$, where $\alpha$=x,y,z labels the components of the current, and $\hat{\mathbf{1}}$ is a unit vector in the direction of the (anti-)clockwise motion of the hole around the impurity. [3]

In the presence of the boundary condition imposed by the skyrmionic configuration of the far field, $S^z_{tot}$ is no longer a good quantum number. As shown in Fig. 2(a) for the parameter values J/t=0.4, V/t=0.5, and t=1.0, $S^z_{tot}$ is a function of $\lambda$ which changes its sign around $\lambda_{opt} \approx 2$ lattice spacings. In the texture form (Eq. (3), the sign of the sublattice magnetization is fixed at large distances ($r \gg \lambda$), and the z-component of the exchange field at the boundary of the inner cluster is either parallel ($\lambda \ll$R) or antiparallel ($\lambda \gg$R) to this magnetization, leading to a net $S^z_{tot} = \pm\frac{1}{2}$ in the two limiting cases,$\lambda \to 0$ and $\lambda \to \infty$, as seen in Fig. 2(a).

A transition is also observed in the z-component of the associated current defined above. In Fig. 2(b) the current along the 8 nearest neighbor sites of the $Li^+$ ion is shown as a function of the far field boundary condition. At $\lambda_{opt}$, it jumps from $\approx$+0.35t to $\approx$ -0.35t. Correspondingly, the total energy of the inner cluster, $E_{cluster}$, shown in Fig. 2(c), has a minimum at $\lambda_{opt}$. Note that $E_{cluster}(\lambda \to 0) = E_{cluster}(\lambda \to \infty)$ because of time-reversal symmetry. While the spin contribution to the total energy of the cluster is largest around $\lambda_{opt}$, the magnitude of the Coulomb energy - which is about an order of magnitude smaller than the spin contribution - is maximal at $\lambda = 0$ and $\lambda = \infty$, i.e. when the mobility of the hole is suppressed by the far field, and it is forced into the vicinity of the impurity.

Special care has to be taken when including renormalization effects due to quantum fluctuations in the spin background. [4] In particular, the $\approx 40\%$ reduction of the sublattice magnetization leads to a reduction of the classical spin stiffness $\rho_s^2 = \mathrm{J} \cdot \mathrm{S}^2$ roughly by a factor of 2. [10] A corresponding reduction also enters the applied spin field at the boundaries of the inner cluster.

The energy of the skyrmion far field is given by

$$\mathrm{E}_{\mathrm{far}} = \frac{\rho_s^2 4\pi\lambda^2}{\mathrm{r}_0^2 + \lambda^2}, \quad (5)$$

where $\mathrm{r}_0(\approx \sqrt{5})$ is the equivalent "radius" of the finite cluster which is treated exactly. $\mathrm{E}_{\mathrm{far}}$ is thus a monotonically increasing function of $\lambda$. The total energy of the system , including the inner cluster and the far field, is shown in Fig. 2(d). The effect of $\mathrm{E}_{\mathrm{far}}$ is a slight reduction of the optimal skyrmion core radius $\lambda_{opt}$.

From Fig. 2, it is observed that the system favors a skyrmion texture with a core size of 2-3 lattice spacings. The results shown in this graph are found to be robust for a wide range of parameters (J,t,V), [11] leading us to the conclusion that a skyrmionic spin configuration is favored for the lightly Li-doped system.

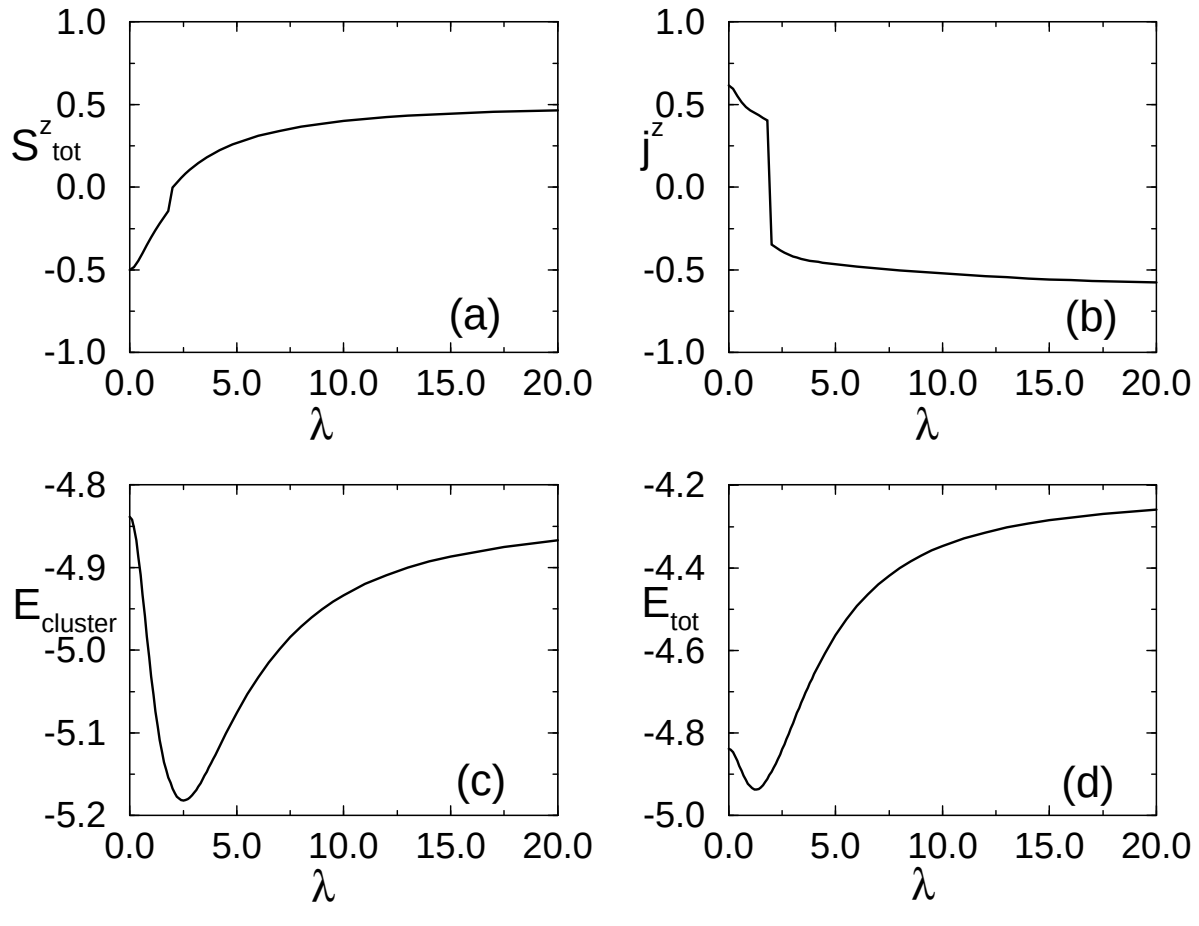

FIG. 2. (a) Total magnetization, $\mathrm{S}^z_{\mathrm{tot}}$, as a function of the skyrmion core size $\lambda$. (b) z-component of the current in the two innermost shells around the $\mathrm{Li}^+$ ion. (c) Energy (in units of t) of the inner cluster (Fig. 1), including hopping, spin and Coulomb contributions. (d) Total energy, including the contribution from the far field (Eq. (5)).

Possible experimental probes of a skyrmion ground state should examine its magnetic nature. The Fourier-transform of the staggered magnetization and the magnetic structure factor exhibit a less sharply peaked behavior in k-space than ordered antiferromagnets. To quantify this statement, here we calculate these quantities for a small concentration of impurities, neglecting skyrmion-skyrmion interactions. [12]

The Fourier-transform of the staggered magnetization, $\mathrm{S}^z_{\mathbf{r}} = (-1)^{x+y}\frac{1}{2}\frac{r^2-\lambda^2}{r^2+\lambda^2} = (-1)^{x+y}\frac{1}{2}[1 - \frac{2\lambda^2}{r^2+\lambda^2}]$, consists of a $\delta$-function at $\mathbf{Q} = (\pi,\pi)$ corresponding to the AF order, and a *tail* due to the topological texture, $\frac{1}{2}[\delta_{\mathbf{q},\mathbf{Q}} - \frac{4\pi\lambda^2}{N}\mathrm{K}_0(\lambda|\mathbf{q}-\mathbf{Q}|)]$, where $\mathrm{K}_0(x)$ is the $0^{\mathrm{th}}$ order modified Bessel function of the second kind. Hence, for a dilute concentration of skyrmions (with concentration $\Gamma_{\mathrm{imp}}$), a deviation from the AF Fourier-transform of the staggered magnetization of the form $\Gamma_{\mathrm{imp}} 2\pi\lambda^2\mathrm{K}_0(\lambda|\mathbf{q}-\mathbf{Q}|)$ is expected.

Similarily, elastic neutron scattering experiments, measuring the magnetic structure factor, $\mathbf{S}^z(\mathbf{q}) = \frac{1}{N}\sum_{\mathbf{r},\mathbf{r}'} e^{i\mathbf{q}\cdot(\mathbf{r}-\mathbf{r}')}\langle \mathbf{S}^z_{\mathbf{r}}\mathbf{S}^z_{\mathbf{r}'}\rangle$, should also pick up a "skyrmion tail",

$$\mathbf{S}^z(\mathbf{q}) = \frac{1}{4}[\delta_{\mathbf{q},\mathbf{Q}} - \Gamma_{\mathrm{imp}} 4\pi\lambda^2\mathrm{K}_0(\lambda|\mathbf{q}-\mathbf{Q}|)] + \mathrm{O}(\Gamma_{\mathrm{imp}}^2\lambda^4). \quad (6)$$

In Fig. 3, the magnetic structure factor is plotted for various values of $\lambda$ at a fixed impurity concentration $\Gamma_{\mathrm{imp}} = 0.2$ %.

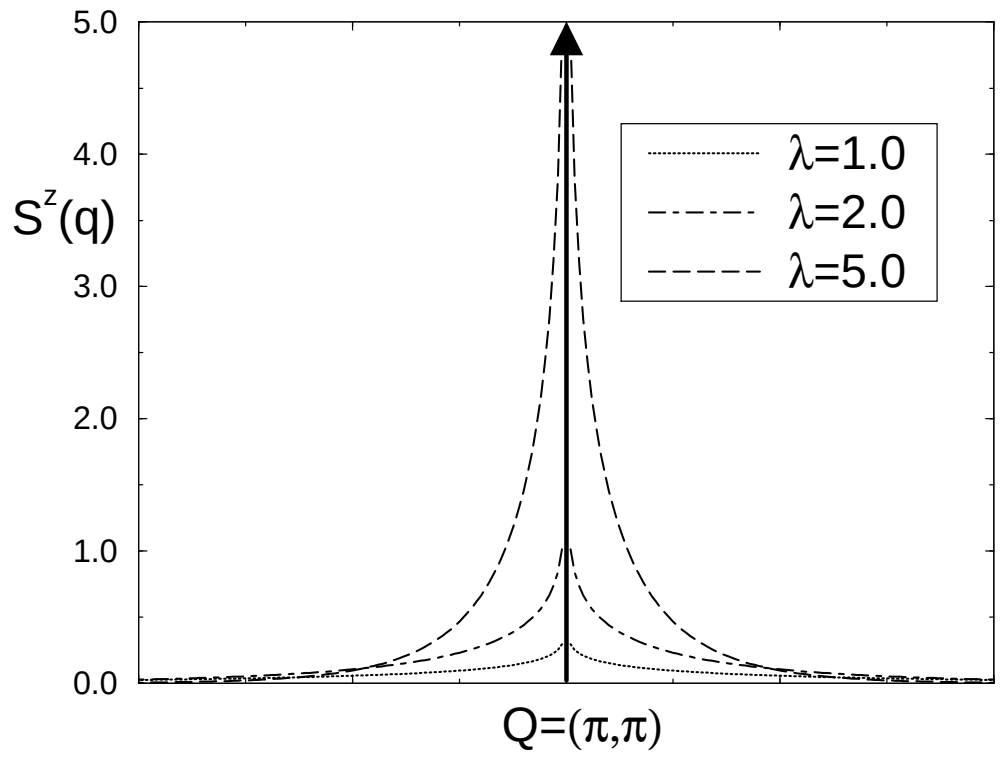

FIG. 3. Magnetic structure factor in the presence of a dilute concentration of skyrmions with core size $\lambda$, for an impurity concentration $\Gamma_{\mathrm{imp}} = 0.2$ %. A deviation from the AF signal ($\delta$-function at $\mathbf{Q} = (\pi,\pi)$) in the form of tails is observed.

A skyrmion spin texture should also have an observable effect on NMR lineshapes. NMR measurements probe the response of the anisotropic hyperfine coupling to local $\mathrm{Cu}^{2+}$ electronic spins and the isotropic transferred hyperfine coupling to the neighboring $\mathrm{Cu}^{2+}$ spins. [13] If $\mathbf{\Phi}_i = \mathrm{A}_x\mathrm{S}^x_i\hat{\mathbf{x}} + \mathrm{A}_y\mathrm{S}^y_i\hat{\mathbf{y}} + \mathrm{A}_z\mathrm{S}^z_i\hat{\mathbf{z}} + \mathrm{B}\sum_{j\in\{i\}}\mathbf{S}_j$ is the hyperfine field, then $\mathrm{I}(\omega) = \sum_i \delta(\omega - |\mathbf{\Phi}_i|^2)$ gives the lineshape which can be observed in measurements of the zero field shift. In the presence of a small concentration of skyrmions this response function can be calculated from Eq. (3), assuming that at large distances from the impurities the spins align antiferromagnetically within the plane, i.e. $\hat{\mathbf{z}} \parallel \hat{\mathbf{b}}$, where $\hat{\mathbf{b}}$ is one of the in-plane axes. Results for various small Li-concentrations using typical values of the hyperfine couplings ($A_x = A_y = 0, A_z = -4B$) are shown in Fig. 4. While at concentrations $\leq 0.06\%$ the lineshape is observed to be rather sharply peaked at $\omega_0$, the resonance frequency for AF order, a shift of the peak towards higher frequencies, along with an asymmetric broadening, occurs at doping levels $\geq 0.2\%$. From the corresponding continuum model it can be shown that the high energy cut-off for the zero field shift NMR signal is at $2\omega_0$.

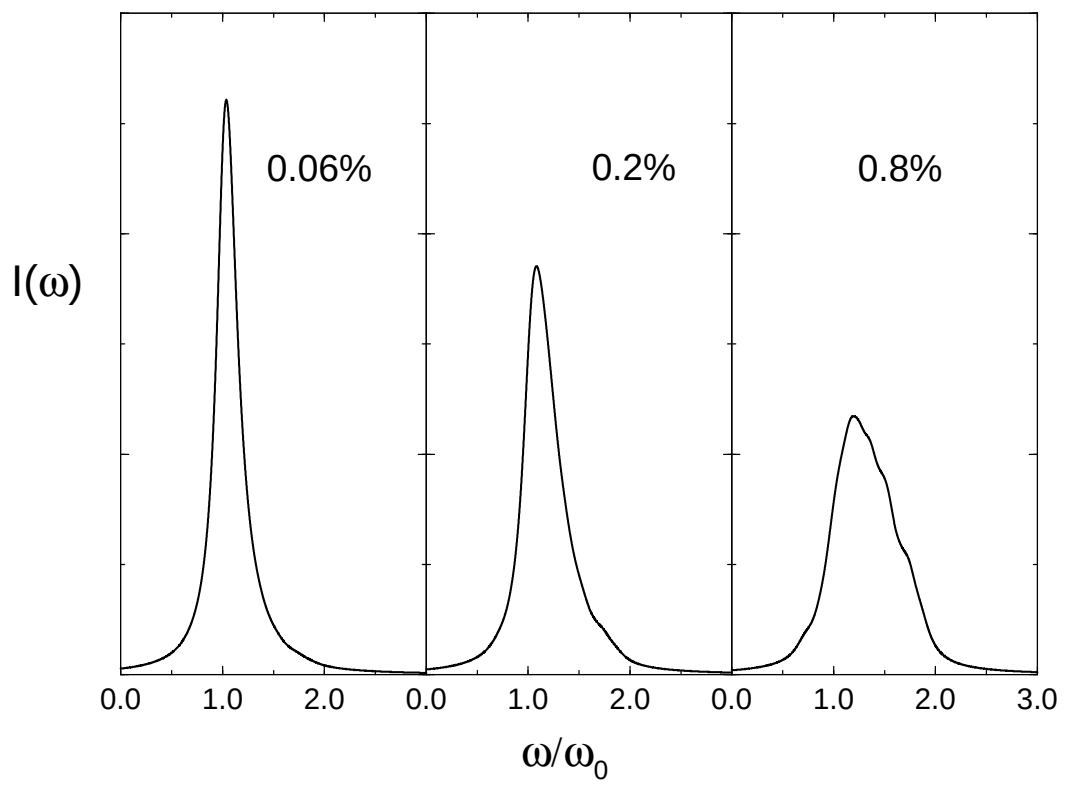

FIG. 4. Zero field NMR lineshape as a function of Li-concentration. The poles appearing in the response function have been given a finite width of $\epsilon = 0.05\omega_0$.

From susceptibility measurements on impurity-doped cuprates it is known that the AF phase is destroyed by Li-substitution of approximately 3%, while Zn-substitution of order 30% is required to suppress AF. [1] Neither of these in-plane substitutions leads to a metallic ground state. In the case of a dilute concentration of $Zn^{2+}$ substitutions for $Cu^{2+}$, no holes are introduced, and the spin background is only locally distorted by the presence of inert sites. $Li^+$ substitution, on the other hand, brings holes into the $CuO_2$ planes, yielding a long ranged distortion of the spin background (Eq. (3)). The holes are localized around the $Li^+$ ions due to attraction via Coulomb forces, since $Li^+$ has a net negative charge with respect to $Cu^{2+}$. [6] Out-of-plane $Sr^{2+}$ substitutions for $La^{3+}$ also attract holes via Coulomb interaction leading to a similar scenario for the spin texture. [2,3,14] However, from transport measurements it is known that a $Sr^{2+}$ concentration of about 3% is sufficient to produce a metallic ground state. Thus we may conclude that in-plane holes are better localized by in-plane $Li^+$ than by out-of plane $Sr^{2+}$.

In conclusion, we have studied the effect of a small number of in-plane $Li^+$ ions on the charge and spin structure in $CuO_2$ planes. Holes are found to be localized around these impurities, to which they are attracted via broken-bond minimization and Coulomb forces. Thus the local magnetic moments on $Cu^{2+}$-sites in the vicinity of the defect are reduced. A dilute concentration of Li-impurities is shown to favor a skyrmion spin texture, leading to "tails" in the static magnetic structure factor [12] and to a broad and asymmetric lineshape of the zero field NMR shift.

We wish to thank J. Sarrao, B. Normand, T. Imai, and M. Sigrist for useful discussions, and acknowledge the Swiss National Science Foundation for financial support.

---